\documentclass[nohyper,12pt,letterpaper]{JHEP3}
 \usepackage{graphicx}
\usepackage{dsfont}
\usepackage{axodraw}

\def\centeron#1#2{{\setbox0=\hbox{#1}\setbox1=\hbox{#2}\ifdim
\wd1>\wd0\kern.5\wd1\kern-.5\wd0\fi
\copy0\kern-.5\wd0\kern-.5\wd1\copy1\ifdim\wd0>\wd1
\kern.5\wd0\kern-.5\wd1\fi}}
\def\ltap{\;\centeron{\raise.35ex\hbox{$<$}}{\lower.65ex\hbox{$\sim$}}\;}
\def\gtap{\;\centeron{\raise.35ex\hbox{$>$}}{\lower.65ex\hbox{$\sim$}}\;}

\def\lsim{\mathrel{\ltap}}
\def\lsup#1{^{\lower 4pt\hbox{$\scriptstyle#1$}}}

 \def\fig#1{Fig.~\ref{#1}}

\def\eq#1{eq.~(\ref{#1})}
 
\def\ddel{\!\!\mathrel{\raise1.5ex\hbox{$\leftrightarrow$\kern-.85em
\lower1.7ex\hbox{$\partial$}}}}

 \title{Note on the pseudo-Nambu-Goldstone Boson of Meta-stable SUSY Violation}
 \author{Tom Banks\\
 Department of Physics and SCIPP\\
 University of California, Santa Cruz, CA 95064\\
 E-mail: \email{banks@scipp.ucsc.edu}\\
 {\it and}\\
 Department of Physics and NHETC, Rutgers University\\
 Piscataway, NJ 08540}

 \author{Howard E.~Haber\\
 Department of Physics and SCIPP\\
 University of California, Santa Cruz, CA 95064\\
 E-mail: \email{haber@scipp.ucsc.edu}\\
 }

 \abstract{ Many models of meta-stable supersymmetry (SUSY) breaking
   lead to a very light scalar pseudo-Nambu Goldstone boson (PNGB),
   $\mathcal{P}$, associated with spontaneous breakdown of a baryon
   number like symmetry in the hidden sector.  Current particle
   physics data provide no useful constraints on the existence of
   $\mathcal{P}$.  For example, the predicted decay rates for $K
   \rightarrow \pi + \mathcal{P}$, $b\to s+\mathcal{P}$ and
    and $\Upsilon\to\gamma+\mathcal{P}$
   are many orders of magnitude below the present experimental bounds.
   We also consider astrophysical implications of the PNGB and find a
   significant constraint from its effect on the evolution of red
   giants. This constraint either rules out models with a hidden
   sector gauge group larger than SU(4), or requires a new
   intermediate scale, of order at most $10^{10}$ GeV, at which the
   hidden sector baryon number is explicitly broken.}

 \received{} \accepted{}
 \preprint{SCIPP-09/13{}\\
August, 2009\\ 
arXiv:0908.2004v3 [hep-ph]\\}
\begin{document}

\section{Meta-stable SUSY breaking and PNGBs}

The idea that, within the quantum field theory approximation,
supersymmetry (SUSY) is broken in a meta-stable vacuum, has a very
long history (for a comprehensive set of references,
see e.g., \cite{iss} and \cite{R}).
 It is however only rather recently, that
this idea has been pursued with vigor. To a large extent, this is
due to the ground-breaking paper of Intriligator, Seiberg and
Shih \cite{iss}, which demonstrated that a large class of
vector-like SUSY gauge theories (based on classical gauge groups) possess
meta-stable SUSY violating vacua. These authors also argued that
field-theoretic meta-stability was more or less {\it required} by a
combination of theoretical and phenomenological arguments \cite{R}.

One of the striking features of these models is that the meta-stable
vacuum breaks the analog of baryon number in the hidden sector (we
will call this meta-baryon number, after the meta-stable state in
the hidden sector), giving rise to a Nambu-Goldstone boson. General
models of this type give rise to other pseudo-Nambu-Goldstone bosons, many
of which would be ruled out by experiment if they were light. This
can be avoided by a variety of methods, but it is interesting that
the models based on SUSY QCD with $N_F = N_C$ have only the
universal penta-baryon pseudo-Nambu-Goldstone boson.
For $N_F = N_C =5$, that particle was
called the penton in \cite{pentagon}.  In this paper, we shall be
more general and consider the model-independent properties of the universal
pseudo-Nambu-Goldstone boson (henceforth designated by PNGB),
and determine constraints on models that are imposed by
current experimental bounds on light CP-odd scalars.

\begin{figure}[t!]
\begin{center}
\begin{picture}(200,50)(0,0)
\DashLine(0,0)(60,0){5}
\Gluon(60,0)(170,30){4}{5}
\Gluon(60,0)(170,0){4}{5}
\Gluon(60,0)(170,-30){4}{5}
\ArrowLine(170,30)(220,0)
\ArrowLine(220,0)(170,-30)
\ArrowLine(170,-30)(170,30)
\Vertex(60,0){5}
\Vertex(220,0){3}
\Text(-20,0)[]{$\partial_\mu \mathcal{P}$}
\Text(178,0)[]{$q$}
\Text(195,29)[]{$q$}
\Text(195,-29)[]{$q$}
\Text(260,0)[]{$\overline q\gamma^\mu q$}
\end{picture}
\end{center}
\vspace{0.3in}
\caption{Diagrams giving rise to a dimension-five coupling
of the PNGB, $\mathcal{P}$.  The PNGB couples to colored
hidden sector quark and squark fields, which generates an
effective operator (indicated by the darkened circle) in which
$\mathcal{P}$ is derivatively-coupled to gluons.  The gluons couple
to hadronic flavor-neutral vector currents of the Standard Model
(through its couplings to the quarks $q$).
\label{dim5}}
\end{figure}
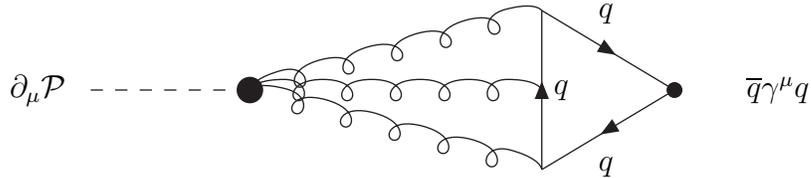

In \cite{pentagon}, one of the present authors argued that the dominant
coupling of the PNGB to Standard Model fields arises from its coupling
to hidden sector quarks and squarks that transform non-trivially with
respect to the ordinary color SU(3) gauge group of the Standard Model.
This is generic in models of direct gauge
mediation~\cite{direct}, in which an SU(3)$\times$SU(2)$\times$U(1)
subgroup of the global flavor symmetry group of the hidden sector
superfields is gauged and identified as the Standard Model gauge
group.  At low-energies, an effective operator arises in which the
PNGB is derivatively-coupled to gluons, as shown in \fig{dim5}.  This
diagram yields a local interaction of the PNGB (henceforth denoted by
$\mathcal{P}$) to neutral flavor hadronic currents of Standard Model,
denoted below by $F^\mu$,
\begin{equation} \label{eq1}
\mathcal{L}_{\rm int}\sim\alpha_3^3(\Lambda_{\rm ISS})
\frac{\Delta m_q^2}{\Lambda_{\rm ISS}^2}
\frac{F^\mu}{\Lambda_{\rm ISS}}\partial_\mu \mathcal{P}
\,.
\end{equation}
For example, $F^{\mu}$ can be any one of the neutral flavor currents, $I_3$,
strangeness, charm, truth or beauty and $\Delta m_q$ is an
appropriate quark mass difference. Integrating by parts, we see that
the dominant coupling comes from the breaking of flavor symmetries
by the charged current weak interactions. The claim in
\cite{pentagon} was that the PNGB would be produced primarily
in association with a conventional flavor changing weak decay.

A brief explanation is in order as to why the effective coupling of
the PNGB involves \textit{three} gluons,\footnote{This result
corrects the previous estimates of the coefficient of the PNGB
interactions with neutral
flavor currents of the Standard Model given in \cite{pentagon}.}
to leading order in the QCD
coupling, $\alpha_3$.  In
typical models of the hidden sector, the theory of the hidden sector
quarks and squarks is vector-like (hence the meta-baryon number
symmetry is non-anomalous) and separately conserves the discrete
symmetries C, P and T. The meta-baryon current, $J^\mu_{\rm MB}$ is
a vector current, which can create a PNGB from the vacuum,
\begin{equation} \label{vacuum}
\langle{0|J^\mu_{MB}(0)|\mathcal{P}}\rangle=f_{\mathcal{P}} q^\mu\,,
\end{equation}
where $q^\mu$ is the PNGB four-momentum and $f_{\mathcal{P}}$ is the
PNGB decay constant.  It follows from \eq{vacuum}
that C$(\mathcal{P})=-1$ and P$(\mathcal{P})=+1$.    That is,
$\mathcal{P}$ is a CP-odd, C-odd scalar.  Color and C conservation
then imply that the minimum number of gluons that the PNGB can
couple to is three.  Remarkably, the form of the P and C-conserving
invariant amplitude for the coupling of a CP-odd, C-odd scalar
to photons was first obtained
by Dolgov and Ponomarev in 1967~\cite{dolgovP}, and subsequently rewritten
as an effective Lagrangian by Dolgov~\cite{dolgov}.
Generalizing the latter result to
gluons, the relevant C and P-invariant
effective Lagrangian with the least number of
derivatives is unique:
\begin{equation} \label{GGG}
\mathcal{L}_{\rm eff}=\frac{g_3^3}{\Lambda_{\rm ISS}^6}d_{abc}
(D_\rho G_{\alpha\beta})^a (D^\beta G_{\sigma\tau})^b
(D^\rho D^\alpha G^{\sigma\tau})^c\, \mathcal{P}\,,
\end{equation}
where $(D_\rho
G_{\alpha\beta})^a\equiv D_{\rho}^{ab}G_{\alpha\beta}^b$, etc.
In \eq{GGG},
$G^a_{\mu\nu}\equiv \partial^\mu A_\nu^a-\partial^\nu A_\mu^a
-gf_{abc}A_\mu^a A_\nu^b$ is the gluon field strength tensor,
$A_\mu^a$ is the gluon field,
$D_\mu^{ab}\equiv\delta^{ab}\partial_\mu+gf_{abc}A_\mu^c$ is
the covariant derivative acting on an adjoint field, and
$d_{abc}\equiv 2{\rm Tr}(\{T_a,T_b\}T_c)$ is the totally symmetric
tensor of color SU(3).

In \fig{dim5}, the dominant contribution to the graph arises when
the internal gauge boson lines are significantly off-shell
and carry momenta of order $\Lambda_{\rm ISS}$.  As a result, the
$\Lambda_{\rm ISS}$ dependence in \eq{eq1} is simply a consequence of
dimensional analysis.  In contrast, in cases where the internal
gauge bosons carry low momenta (which for gluons would lead to
strong coupling), one should perform an operator product expansion first
in the gauge theory coupled only to the hidden sector and then
consider the effect of those pure gauge operators in the Standard
Model.  The leading gluonic operator is given in \eq{GGG}, and the
resulting effective
coupling is highly suppressed for momenta small compared
to $\Lambda_{\rm ISS}$.  Similarly, the decay
of the PNGB into three photons is governed by \eq{GGG} with
$d_{abc}=1$, $g_3$ replaced by $e$, $D_\mu$ replaced by $\partial_\mu$, and
$G^a_{\mu\nu}$ replaced by the electromagnetic field strength $F_{\mu\nu}$.
Due to the $\Lambda_{\rm ISS}^{-6}$ suppression in \eq{GGG}, the
decay rate of the PNGB into three photons is many orders of magnitude
smaller than the inverse lifetime of the universe, and poses no
significant cosmological constraint.

In this paper we will examine more closely the properties
of the PNGB in order to determine whether the experimental
non-observation of a light CP-odd particle can impose significant
constraints on model building.  In Section 2, we will estimate the
mass of the PNGB.  To the extent that the global flavor symmetries of
the hidden-sector Lagrangian are exact, the PNGB would be an exactly
massless Nambu-Goldstone boson.  However, in realistic models, we
expect a small violation of these flavor symmetries arising at energy
scales much higher than $\Lambda_{\rm ISS}$.  Taking such explicit
breakings into account implies that the Nambu-Goldstone boson is
actually a \textit{pseudo}-Nambu-Goldstone boson with a small mass
proportional to the relevant symmetry-breaking parameter.

In Sections 3 and 4, we will sharpen the estimates of the coefficients of the
effective interactions of the PNGB with Standard Model particles.
We will clarify the energy scales involved in determining the
relevant effective operators.
We also note that there is an important new process which is made
possible by this interaction; the direct conversion of the hadronic
PNGBs (namely, the pions and kaons)
into each other, with the emission of a PNGB. Bounds on
such processes have been investigated in connection with the idea of
``flavor Goldstone bosons''~\cite{FN}.  The experimental constraints on $K$ to
$\pi$ meson conversion are the strongest.  Nevertheless, we
demonstrate that the predicted decay rate for $K\to\pi\mathcal{P}$
is significantly below the present experimental bounds.

When the gauge bosons of \fig{dim5} are
SU(2)$\times$U(1) gauge bosons, then
such operators generate effective
flavor-diagonal and flavor non-diagonal
Yukawa couplings, whose magnitudes can potentially be constrained by
experimental data.  For example, the former provides a mechanism for the decay
$\Upsilon\to\gamma\mathcal{P}$, whereas the latter provides the dominant
contribution to  $K\to\pi\mathcal{P}$ and $b\to s\mathcal{P}$.  
In all cases, the predicted
decay rates are many orders of magnitude below the
corresponding experimental bounds.
In contrast,
astrophysical consequences of the PNGB investigated in Section~5 imply
that its effective Yukawa coupling to electrons is larger than that
allowed by bounds on stellar cooling. Thus, one must arrange that
its mass be large enough, such that it cannot be produced in stars. We
present a preliminary analysis of the model building constraints
imposed by this bound.

Although our analysis was motivated by the PNGB of models of
meta-stable SUSY breaking, it is actually much more general.
It applies to any hidden sector theory
whose characteristic energy scale is in the TeV to multi-TeV range (the
scale enters only into the explicit numerical estimates), which
produces a light CP-odd, C-odd spin zero PNGB.
 In this connection, it is
worth noting that the famous Vafa-Witten result~\cite{Vafa:1983tf},
that vector
symmetries are not spontaneously broken in QCD-like gauge theories
does {\it not} apply to supersymmetric theories.  The positivity of
the Dirac determinant, crucial to the Vafa-Witten argument, is
violated by the Yukawa couplings in the presence of scalar
backgrounds. Thus the existence of CP-odd, C-odd PNGBs in SUSY
theories might be much more general than a particular class of
models.  Our results can be viewed as the first step in a general
analysis of the properties of such particles.

Finally, we note that another paper on hidden sector Nambu-Goldstone
bosons \cite{dedes} has appeared recently.  These authors make
different assumptions about the leading coupling of the PNGB to the
Standard Model, from those that are natural in models of
meta-stable SUSY breaking.   There is little overlap between our
analysis and theirs.

\bigskip\bigskip
\section{The PNGB mass}

Meta-baryon number is an anomaly free global accidental symmetry of
all models based on the ISS mechanism of meta-stable SUSY
breaking, and is spontaneously broken in the meta-stable vacuum. In
this approximation, the PNGB is an exact Nambu-Goldstone
boson. However, we expect that there are irrelevant corrections to
the effective Lagrangian that explicitly break this symmetry.  In a
model with SU$(N_C)$ color, the lowest dimension gauge invariant
operator carrying meta-baryon number, has dimension $N_C$.  This is
a chiral operator, and contributes terms of dimension $\geq N_C + 1$
to the Lagrangian. For example,
in the Pentagon model, $N_C = 5$ but there is an
R symmetry, under which the penta-baryon operators have R charge 0,
so the lowest dimension allowed operator is dimension $7$. Note that
for $N_C = 2,3$, meta-baryon number violation is relevant or
marginal and there is no PNGB in the spectrum.\footnote{Actually, this
depends on how the Standard Model fits into the flavor group
SU$(N_F)$. There may be no components of the meta-baryon that are
Standard Model singlets.}  In this paper, we will assume $N_C \geq
4$.

We will also assume that the high energy scale associated with these
operators is the unification scale, $M_U = 2 \times 10^{16}$ GeV.
The reader can easily modify our results to replace this by the reduced
Planck scale, $m_P = 2 \times 10^{18}$ GeV, or the apparent neutrino
see-saw scale $M_{SS} \sim 5 \times 10^{14}$ GeV (which is a
dangerous scale for ordinary dimension $6$ baryon violating
operators).
The lowest gauge-invariant operator (involving fields of the electric theory)
that violates meta-baryon number, which contributes to the hidden
sector superpotential, is given by
$$
\delta W\sim\frac{1}{\Lambda_U^{N_c-3}} \mathcal{Q}^{N_c}\,,
$$
where $\mathcal{Q}$ is a hidden sector quark superfield.
If the above operator is disallowed (due, say,
to discrete symmetries preserved at the scale $M_U$), then one can introduce
an extra singlet field $S$ and choose,
$$
\delta W\sim\frac{1}{\Lambda_U^{N_c+P-3}} \mathcal{Q}^{N_c}S^P\,,
$$
for some suitably chosen $P$.  In either case, the PNGB, $\mathcal{P}$,
acquires a non-trivial potential due to the explicit breaking,
$$ V =
\Lambda_{\rm ISS}^4 \left({{\Lambda_{\rm ISS}}\over M_U}\right)^{N_C + P -3}
U(\mathcal{P}/\Lambda_{\rm ISS} ),$$
where $U(x)=U_0+cx^2+\cdots$, for some constant $c\sim\mathcal{O}(1)$.

The PNGB mass is then
$$m_{\mathcal{P}} \sim \Lambda_{\rm ISS}
\left({{\Lambda_{\rm ISS}}\over M_U}\right)^{{N_C + P
-3 } \over 2} ,$$
The choice of $M_{\rm ISS}$ and $M_U$ is highly model-dependent.
In the framework of gauge-mediated SUSY-breaking models,
one expects $\Lambda_{\rm ISS}$ to be in the TeV to multi-TeV range.
In this work we choose:
$$
\Lambda_{\rm ISS}\sim 2~{\rm TeV}\,,
$$
which is an optimistic choice (most probably this scale is
significantly larger).  If we allow a possible range of unification
masses, $5\times 10^{14}$~GeV$\lsim M_U\lsim 2\times 10^{18}$~GeV, as
indicated above and consider three possible values of $N_C+P=4,5,6$,
then we find possible PNGB masses lying in the range:
$$
6\times 10^{-11}~{\rm eV}\lsim m_{\mathcal{P}}\lsim 4~{\rm MeV}\,,
$$
which is a huge dynamic range of possible masses.

In \cite{pentabaryogen}, a version of the Pentagon model in which the
PNGB accounted for both baryogenesis and dark matter requires a
meta-baryon number breaking scale of order $10^8$---$10^{10}$ GeV in
place of $M_U$.  In this latter case, the
corresponding PNGB mass is given by
$m_{\mathcal{P}} \sim 100~\hbox{\textrm{MeV---100~keV}}.$

\section{PNGB interactions via QCD interactions}

The dominant coupling of PNGBs to the Standard Model comes, at
lowest order in Standard Model couplings, through the diagrams of
\fig{dim5}. We can split this diagram, and QCD corrections to it,
into pieces where the gluon lines carry momentum of order
$\Lambda_{\rm ISS}$ and contributions from lower scales.  The form of
the lower scale contributions involves an operator expansion in
gauge invariant pure glue operators, which is then inserted into low
energy QCD.  All higher order QCD corrections must be included in
such contributions.

The $\mathcal{P}$ field is a Nambu-Goldstone boson, so the resulting effective
action is a sum of contributions of the form
$$g_3^3 (\Lambda_{\rm ISS} )\partial_{\mu} \mathcal{P} {{V^{\mu}} \over
{\Lambda^{D-2}_{\rm ISS}}}, $$ where $V^{\mu}$ is a gauge invariant,
pure glue operator of dimension $D$.  In QCD, if we assume the
hidden sector conserves P and C, then $D \geq 8$ for such
operators. There are also terms higher order in the $\mathcal{P}$
field, or with higher derivatives, that are even more suppressed.

The high momentum part of the one-loop (QCD) diagram, gives
interactions of the form
\begin{equation} \label{highmom}
\alpha_3^3 (\Lambda_{\rm ISS} ) \partial_{\mu} \mathcal{P} {{F^{\mu}} \over
\Lambda_{\rm ISS}} ,
\end{equation}
where $F^{\mu}$ is some neutral hadronic vector
current.\footnote{In order to conserve C, the hadronic
current $F^\mu$ must be a vector
current and not a pseudo-vector current.}
The third power of the QCD coupling
is a consequence of C-invariance which requires at least three gluons
in \fig{dim5}.  It is convenient to decompose $F^\mu$ into a sum of
the ordinary baryon number current,
$J_B^\mu\equiv\sum_i \bar{Q}_i\gamma^\mu Q_i$ and
\begin{equation} \label{fmu}
F^\mu\sim \bar{Q}\gamma^\mu T^a Q\,,\qquad {\rm where}~~{\rm Tr}~T^a=0\,,
\end{equation}
Consider first $F^\mu$ given by \eq{fmu}.
The dominant momentum passing through \fig{dim5} is of
$\mathcal{O}(\Lambda_{\rm ISS})$.
Thus, we can treat the quark propagators in the mass insertion approximation.
An even number of mass insertions is required.  The triangle
with no mass insertions vanishes, since ${\rm Tr}~T^a=0$.  Thus, the
first non-vanishing result derives from the diagram
with two mass insertions and yields a result
proportional to ${\rm Tr}(M^2 T^a)$.  The latter vanishes for
degenerate quark masses; hence the result given in \eq{highmom}
must be suppressed further by a factor of
$$ {{\Delta m^2_q}\over
\Lambda^2_{\rm ISS}},$$ where $\Delta m^2_q$ is an appropriate quark
squared-mass difference.  Thus, we confirm the estimate given in
\eq{eq1}. Note that in the low momentum part of the
diagram, flavor violation only costs powers of $\Delta m_q/\Lambda_{QCD}$.
Nevertheless, the high dimension required by the pure glue
operators makes these contributions smaller than the ones we have
just estimated.

To proceed further with our analysis, we will use the fact that the
couplings we have calculated are all small, and will be used only in
first order perturbation theory. Thus, we can integrate by parts,
and use the equations of motion of the flavor currents in the
unperturbed Lagrangian to get a non-derivative coupling between the
PNGB and Standard Model fermions,
\begin{equation} \label{nonderiv}
\mathcal{L}_{\rm int}\sim\alpha_s^3(\Lambda_{\rm ISS})
\frac{\Delta m_Q^2}{\Lambda_{\rm ISS}^2}
\,\mathcal{P}\, \frac{\partial_\mu F^\mu}{\Lambda_{\rm ISS}}
\,.
\end{equation}
The correct all-orders
procedure is to perform a field redefinition, which has the form of
a flavor gauge transformation, to eliminate the derivative coupling
in favor of an exponential coupling of the PNGB to flavor
changing operators. This would supplement the couplings obtained by
our method by multi-PNGB operators, which are of higher order.

Using this prescription, we see that the coupling of the PNGB to the baryon
number current, which does not have a quark mass difference
suppression, vanishes for low momentum PNGBs (since the baryon number
current is conserved). In principle it
could give rise to PNGB emission in the scattering of weak
bosons, but apart from that it would correspond to emission only
from virtual weak boson lines in high order weak interactions.

In contrast, the non-singlet flavor current divergences are
predominantly due to first order charged current weak interactions.
As an example, consider the strangeness current $J^\mu_S=\bar s\gamma^\mu s$.
Thus, at energy scales below the mass of the $W$ boson,
$$
\partial_\mu J^\mu_S=\frac{G_F\sin\theta_c}{\sqrt{2}}\left[\bar{u}
\gamma_\nu(1-\gamma_5 )d\,\bar{s}\gamma^\mu(1-\gamma_5)u+{\rm h.c.}\right]\,,
$$
due to the effective $\Delta S=\pm 1$ four-Fermi weak interaction.
Inserting this result into \eq{nonderiv} yields the effective
interaction Lagrangian
$$ \mathcal{L}_{\rm int}=
\alpha_3^3 (\Lambda_{\rm ISS} ) {m^2_s\over \Lambda^3_{\rm ISS}}
{{G_F \sin\theta_c }\over \sqrt{2}} \mathcal{P}
(J_{\mu}^{0-} J^{\mu}_{1+} + {\rm h.c.})\,,
$$
where $m_s$ is the strange quark mass and
$J_{\mu}^{S\pm}$ and $J^{\mu}_{S\pm}$ indicate
the part of the hadronic weak current that changes strangeness by
$\pm 1$ unit.
 To compute the contribution of \fig{dim5} to the decay
rate for $K^\pm\to\pi^\pm \mathcal{P}$, we use
$\langle 0| \bar{u}\gamma_\nu(1-\gamma_5 )d|\pi^{-}\rangle
=if_{\pi}q_\pi^\mu$ and similarly for $K$.  This corresponds to
replacing the two currents by $J_{\mu}^{0-} \rightarrow f_{\pi}
\partial_{\mu} \pi^{-}$ and $J_{\mu}^{1+} \rightarrow f_{K}
\partial_{\mu} K^{+}$.  The invariant amplitude,
evaluated in the kaon rest frame, is then
$$
\mathcal{M}(K^\pm\to\pi^\pm \mathcal{P})=\frac{G_F}{\sqrt{2}}
\frac{\alpha_s^3(\Lambda_{\rm ISS})}{\Lambda_{\rm ISS}^3} m_s^2\sin\theta_c f_\pi f_K \frac{m_K^2}{2}\,.
$$
where we have used $E_\pi\simeq m_K/2$.
A rough estimate of the partial width for this decay then gives
$$\Gamma(K^\pm \rightarrow  \pi^\pm \mathcal{P}) \sim 10^{-48}~{\rm
  GeV}\left(\frac{2~{\rm TeV}}{\Lambda_{\rm
      ISS}}\right)^6$$
We will show in Section 4 that there are other contributions to
the decay rate for $K^\pm \rightarrow  \pi^\pm \mathcal{P}$
that are significantly larger than the one computed above.  These
arise from effective flavor-changing Yukawa couplings that are
generated by purely weak interaction effects.

\section{PNGB interactions via the electroweak interactions}

In Section 3, we discussed the PNGB interactions that arise
due to strong QCD interactions that couple the
Standard Model quark sector and the
hidden sector quarks and squarks.
However, diagrams such as those exhibited in
\fig{dim5} do not yield flavor \textit{diagonal} couplings of the PNGB to the
Standard Model fields in the limit of low-momentum PNGBs.
As previously noted, after integration by parts, the divergence
of the corresponding flavor-diagonal quark current vanishes.

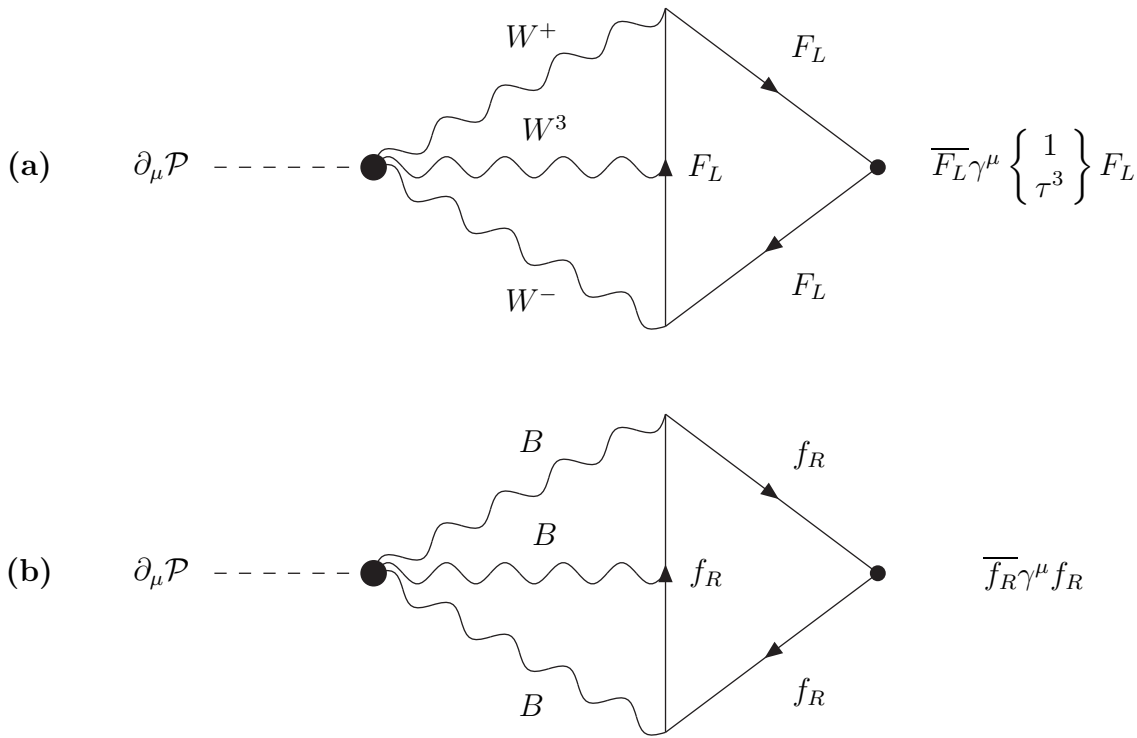
\begin{figure}[b!]
\begin{center}
\begin{picture}(200,50)(30,10)
\DashLine(0,0)(60,0){5}
\Photon(60,0)(170,60){4}{5}
\Photon(60,0)(170,0){4}{5}
\Photon(60,0)(170,-60){4}{5}
\ArrowLine(170,60)(250,0)
\ArrowLine(250,0)(170,-60)
\ArrowLine(170,-60)(170,60)
\Vertex(60,0){5}
\Vertex(250,0){3}
\Text(-20,0)[]{$\partial_\mu \mathcal{P}$}
\Text(225,45)[]{$F_L$}
\Text(186,0)[]{$F_L$}
\Text(225,-45)[]{$F_L$}
\Text(120,50)[]{$W^+$}
\Text(125,15)[]{$W^3$}
\Text(120,-50)[]{$W^-$}
\Text(310,0)[]{$\overline{F_L}\gamma^\mu\left\{\begin{array}{c} 1 \\ \tau^3
\end{array}\right\} F_L$}
\Text(-70,0)[]{\bf (a)}
\end{picture}
\end{center}
\vspace{1.2in}
\begin{center}
\begin{picture}(200,50)(30,0)
\DashLine(0,0)(60,0){5}
\Photon(60,0)(170,60){4}{5}
\Photon(60,0)(170,0){4}{5}
\Photon(60,0)(170,-60){4}{5}
\ArrowLine(170,60)(250,0)
\ArrowLine(250,0)(170,-60)
\ArrowLine(170,-60)(170,60)
\Vertex(60,0){5}
\Vertex(250,0){3}
\Text(-20,0)[]{$\partial_\mu \mathcal{P}$}
\Text(225,45)[]{$f_R$}
\Text(186,0)[]{$f_R$}
\Text(225,-45)[]{$f_R$}
\Text(120,50)[]{$B$}
\Text(125,15)[]{$B$}
\Text(120,-50)[]{$B$}
\Text(310,0)[]{$\overline{f_R}\gamma^\mu f_R$}
\Text(-70,0)[]{\bf (b)}
\end{picture}
\end{center}
\vspace{0.7in}
\caption{Typical electroweak contributions to the coupling of the PNGB
to Standard
Model fermion currents, where
$F_L\equiv (u_L\,,\,\,d_L)$ in the case of a left-handed quark doublet,
$F_L\equiv (\nu_L\,,\,\,e_L)$ in the case of a left-handed lepton doublet,
and $f_R$ is a right-handed quark or charged lepton singlet (with
generation indices suppressed in all cases).
In (a), couplings to both the singlet and triplet currents can
appear.
\label{weakboxes}}
\end{figure}

However, flavor-diagonal couplings of the PNGB to quarks and leptons
can be generated directly by diagrams involving
electroweak gauge bosons.  Two typical diagrams are
shown in Fig.~\ref{weakboxes}.  In computing the flavor-conserving
couplings, one can neglect the deviation of the diagonal CKM matrix elements
from unity.
Moreover, as $g_2>g_1$, we will keep only
the leading contribution that is proportional to $\alpha_2^3$ [and
neglect, e.g., the contribution of the hypercharge gauge boson ($B$)
exchange graph in Fig.~\ref{weakboxes}(b)].

As in our analysis of the graphs of \fig{dim5}, the dominant momentum
in the loops are of $\mathcal{O}(\Lambda_{\rm ISS})$.  Hence, the
resulting PNGB interaction is governed by an operator of the form:
$$
\frac{\alpha_2^3(\Lambda_{\rm ISS})}
{\Lambda_{\rm ISS}}\overline f\gamma^\mu (1-\gamma_5) f\partial_\mu\mathcal{P}\,.
$$
Integrating by parts and using the free field equations then yields
the desired Yukawa coupling:\footnote{ In a C-conserving theory,
$\mathcal{P}$ is a $J^{PC}=0^{+-}$ scalar.
However, the Yukawa couplings of $\mathcal{P}$ to
Standard Model fermions generated
by electroweak physics necessarily introduces C and P violation.  Consequently
$\mathcal{P}$ behaves as a linear combination of $J^{PC}=0^{+-}\oplus 0^{-+}$
(we neglect small CP-violating effects).  As a result,
$\mathcal{P}$ can now couple
diagonally to a fermion-antifermion pair via the pseudoscalar
$\gamma_5$ coupling.}
\begin{equation} \label{Yuk}
\mathcal{L}_{\mathcal{P}\overline ff}\sim \frac{\alpha_2^3(\Lambda_{\rm ISS})m_f}
{\Lambda_{\rm ISS}}\,i\overline f\gamma_5 f\,\mathcal{P}\,.
\end{equation}
In contrast to the flavor-changing couplings that are suppressed by $1/\Lambda_{\rm ISS}^3$,
the flavor-conserving couplings of $\mathcal{P}$ scale as one inverse power of $\Lambda_{\rm ISS}$
(since no mass insertions on the fermion lines are required).

As an application to the above result, we compute the decay rate for
$\Upsilon\to \gamma\mathcal{P}$.  In general, if
$V$ is a $^3 S_1$ quarkonium bound state of $\overline{Q} Q$
(for a heavy quark $Q=c$ or~$b$), and if
the $\mathcal{P} \overline Q
Q$ coupling is given by
$\lambda_Q m_Q i\overline{Q}\gamma_5 Q\mathcal{P}$, then a tree-level
computation yields~\cite{wilczek}:
$$
\frac{\Gamma(V\to\gamma\mathcal{P})}{\Gamma(V\to e^+e^-)}
=\frac{\lambda_{Q}^2m_V^2}{8\pi\alpha}\,.
$$
According to \eq{Yuk}, $\lambda_Q\sim\alpha_2^3/\Lambda_{\rm ISS}$.
Hence, using BR($\Upsilon\to e^+e^-)=2\%$~\cite{pdg}
and $\alpha_2\sim 0.03$, we find:
$$
{\rm BR}(\Upsilon\to \gamma\mathcal{P})
\sim 2\times 10^{-15}\left(\frac{2~{\rm TeV}}{\Lambda_{\rm ISS}}\right)^2\,,
$$
which is many orders below the experimental bound.  The corresponding
branching ratio
for $\psi\to\gamma\mathcal{P}$ is about a factor of three smaller,
which is again much too small to be experimentally ruled out.

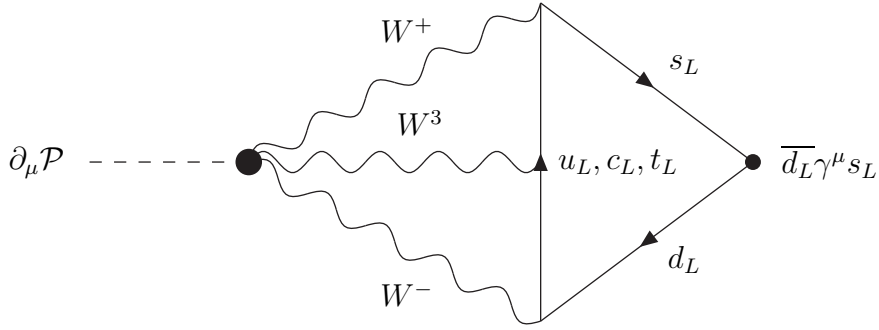
\begin{figure}[ht!]
\begin{center}
\begin{picture}(200,50)(50,0)
\DashLine(0,0)(60,0){5}
\Photon(60,0)(170,60){4}{5}
\Photon(60,0)(170,0){4}{5}
\Photon(60,0)(170,-60){4}{5}
\ArrowLine(170,60)(250,0)
\ArrowLine(250,0)(170,-60)
\ArrowLine(170,-60)(170,60)
\Vertex(60,0){5}
\Vertex(250,0){3}
\Text(-20,0)[]{$\partial_\mu \mathcal{P}$}
\Text(225,37)[]{$s_L$}
\Text(200,0)[]{$u_L, c_L, t_L$}
\Text(225,-37)[]{$d_L$}
\Text(120,50)[]{$W^+$}
\Text(125,15)[]{$W^3$}
\Text(120,-50)[]{$W^-$}
\Text(280,0)[]{$\overline{d_L}\gamma^\mu s_L$}
\end{picture}
\end{center}
\vspace{0.7in}
\caption{A typical electroweak contribution to the off-diagonal
(flavor-changing) couplings of Standard Model fermion currents to the PNGB.
\label{nondiag}}
\end{figure}

Next, we examine the non-diagonal
(flavor-changing) Yukawa coupling generated by the electroweak interactions.
As an example, we compute the effective
$d s\mathcal{P}$ Yukawa interaction, which arises from diagrams such
as the one depicted in \fig{nondiag}.
Inserting the factors of the CKM matrix $V$ at the two charged $W$
vertices of the triangle, we note that if the up-type quark masses
were degenerate one would produce a factor of $V^\dagger
V=\mathds{1}$, resulting in a diagonal coupling.  Thus, in the case of
a non-diagonal coupling, we have a GIM suppression~\cite{GIM}.
Treating the quark
masses in the mass-insertion approximation, one needs two mass
insertions to obtain a flavor-changing coupling.  This yields a
suppression factor of $\Delta m_q^2/\Lambda_{\rm ISS}^2$.  The
contribution from the top
quark in the loop dominates, and the resulting effective
operator for the $ds\mathcal{P}$ interaction is given by:
\begin{equation} \label{dsP}
\frac{\alpha_2^3}{\Lambda_{\rm ISS}^3} m_t^2 V_{td} V^*_{ts}
\overline{d}\gamma^\mu(1-\gamma_5) s\,
\partial_\mu\mathcal{P}+{\rm h.c.}
\end{equation}

In order to compute the decay rate for $K\to\pi\mathcal{P}$,
we employ the techniques of the low-energy chiral Lagrangian.
Following \cite{Georgi}, we identify
\begin{equation} \label{georgi}
\bar{d}\gamma^\mu(1-\gamma_5)s+{\rm h.c.}\longleftrightarrow
-2f_K\partial^\mu {K}_2^0-i\left[K^-\,\ddel\lsup{\,\mu}
\pi^-  -K^+\,\ddel\lsup{\,\mu}\pi^-
+{K}_1^0\,\ddel\lsup{\,\mu}\pi^0\right]+\cdots\,,
\end{equation}
where we have isolated those terms that are linear and quadratic in
the pion and kaon fields.  In the notation above,
$K_1^0\equiv(K^0-\overline{K}\lsup{0})/\sqrt{2}\simeq K^0_S$ is CP-even and 
$K_2^0\equiv(K^0+\overline{K}\lsup{0})/\sqrt{2}\simeq K^0_L$ is CP-odd.  
The matrix element for 
$K^\pm\to\pi^\pm\mathcal{P}$ is thus proportional to $(p_K+p_\pi)\cdot
p_{\mathcal{P}}=m_K^2-m_\pi^2\simeq m_K^2$, and thus we find
\begin{equation} \label{dsp}
\Gamma(K^\pm\to\pi^\pm\mathcal{P})\simeq \frac{\alpha_2^6 m_t^4
|V_{td}V_{ts}^*|^2m_K^3}{16\pi\Lambda_{\rm ISS}^3}\sim 2.5\times 
10^{-30}~{\rm GeV} \left(\frac{2~{\rm TeV}}{\Lambda_{\rm ISS}}\right)^3\,,
\end{equation}
where we have used $\alpha_2^2\sim 10^{-3}$ and $|V_{td} V^*_{ts}|\sim
3\times 10^{-4}$~\cite{pdg}. 

Remarkably, the QCD contribution to the decay rate
for $K^\pm\to\pi^\pm \mathcal{P}$, obtained in
Section 3, is negligible compared with the result of \eq{dsp}.
The $K^\pm$ lifetime is of order $10^{-8} $ sec., corresponding to a
width of about $5\times 10^{-16} $ GeV.
Thus, the branching ratio
for the rare kaon decay into pion plus PNGB is roughly
\begin{equation} \label{rareK}
{\rm BR}(K^\pm\to\pi^\pm\mathcal{P}) \sim 5\times 10^{-15}\left(\frac{2~{\rm
      TeV}}{\Lambda_{\rm ISS}}\right)^6\,.
\end{equation}

Likewise, we can estimate the rate for $K_L^0\to\pi^0\mathcal{P}$.
Using \eq{georgi}, we see that this process is absent in the limit of
CP conservation.  Including the CP-violating effects, we can write
$K_1^0\simeq K_S^0-\epsilon K_L^0$, where $|\epsilon|\sim 2\times
10^{-3}$. Thus, $\Gamma(K_L^0\to\pi^0\mathcal{P})$ is suppressed by
an additional
a factor of $|\epsilon|^2$.  Noting that the $K^0_L$ lifetime is of
order $5\times 10^{-8}$~sec., we estimate
$$
{\rm BR}(K^0_L\to\pi^0\mathcal{P}) \sim 10^{-19}\left(\frac{2~{\rm
      TeV}}{\Lambda_{\rm ISS}}\right)^6\,.
$$

The strongest experimental bounds on such decays \cite{ani} are for
approximately massless PNGBs, and give a branching ratio below
$7 \times 10^{-11}$. For MeV scale masses, the bounds of
\cite{adler} are about three orders of magnitude weaker.  Our
predicted branching ratios are significantly
below these experimental bounds. However, future experiments anticipate
the possibility of detecting 
branching ratios for $K\to\pi+{\rm invisible}$
as small as $10^{-14}$~\cite{projectx},
which is approaching the estimate given in \eq{rareK}.

For the related process $B^\pm\to K^\pm\mathcal{P}$, the chiral
Lagrangian technique is no longer appropriate.  In this case, it is
more useful to consider the inclusive partonic decay process
$b\to s\mathcal{P}$.  In analogy with \eq{dsP}, the effective operator
for the $sb\mathcal{P}$ interaction is also dominated by the top quark
loop and is given by:
$$
\frac{\alpha_2^3}{\Lambda_{\rm ISS}^3} m_t^2 V_{ts} V^*_{tb} 
\overline{s}\gamma^\mu(1-\gamma_5) b\,
\partial_\mu\mathcal{P}\,.
$$
Integrating by parts and using the field equations
yield the desired Yukawa coupling:
$$
\mathcal{L}_{sb\mathcal{P}}\sim 
\frac{\alpha_3^2m_t^2 V_{ts} V^*_{tb}m_b}
{\Lambda_{\rm ISS}^3}(i\overline s\gamma_5 b\,\mathcal{P}+{\rm h.c.})\,.
$$

We now compute the decay width for $b\to s\mathcal{P}$.  The effective
Yukawa coupling is
$$
\lambda\sim \frac{\alpha_2^3m_t^2 V_{ts} V^*_{tb}m_b} {\Lambda_{\rm
    ISS}^3}\sim 2\times 10^{-11}\left(\frac{2~{\rm TeV}}{\Lambda_{\rm
      ISS}}\right)^3\,,
$$
where we have used $|V_{ts} V^*_{tb}|\sim
0.04$ and $m_b=4.2$~GeV~\cite{pdg}.  Thus, 
$$
\Gamma(b\to s\mathcal{P})\simeq\frac{\lambda^2 m_b}{16\pi}\sim 3\times
10^{-23}~{\rm GeV}\left(\frac{2~{\rm TeV}}{\Lambda_{\rm
      ISS}}\right)^6\,.
$$
The $b$ lifetime is roughly $1.6\times 10^{-12}$~sec., corresponding to
a width of about $3\times 10^{-12}$~GeV.
Thus, the inclusive branching ratio for $B$ meson decay into
strange mesons plus a PNGB is roughly
$$
{\rm BR}(b\to s\mathcal{P})\sim 10^{-11}\left(\frac{2~{\rm
      TeV}}{\Lambda_{\rm ISS}}\right)^6\,.
$$
Unfortunately, such a small branching ratio is out of the reach 
of the next generation of super $B$ factories~\cite{bfactory}.

\section{Cosmological and astrophysical effects}

We have been unable to find any cosmological constraints on
the PNGB. For example, if the PNGBs
were the dark matter (e.g., as in \cite{pentabaryogen}), the extra
contributions to proton/neutron conversion due to the interactions
$$n \rightarrow p + e^- + \bar{\nu }_e + \mathcal{P} ,$$ and
especially
$$\mathcal{P} + p \rightarrow n
+ e^+ + \nu_e ,$$ could in principle alter the results of
nucleosynthesis. However, PNGB interactions decouple at a higher
temperature than neutrinos and are found to be
negligible.

Because the neutrino mass is nonzero, it too will have a
Yukawa coupling to $\mathcal{P}$ given by
$$
\lambda_{\nu\nu\mathcal{P}}\sim \frac{\alpha_2^3 m_{\nu}}
{\Lambda_{\rm ISS}}\sim 10^{-19}
\left(\frac{2~{\rm TeV}}{\Lambda_{\rm ISS}}\right)\,.
$$
This leads to a new energy loss mechanism
for supernovae. Neutrinos trapped in the hot plasma can
bremsstrahlung the \textit{very} weakly interacting PNGBs,
which transport energy out of the star. However, the coupling
$\lambda_{\nu\nu\mathcal{P}}$ is too
small for this to be a significant effect.

The coupling of the PNGB to electrons in red giants is a more significant
constraint, because of the marvelous limits on energy loss processes
in these stars.  The electron Yukawa coupling,
$\lambda_{ee\mathcal{P}}$, can be estimating from \eq{Yuk}.  Defining
the ``PNGB fine-structure constant'' by
$\alpha_{\mathcal{P}}\equiv\lambda_{ee\mathcal{P}}^2/4\pi$, it follows
that:
$$
\alpha_{\mathcal{P}}\sim 10^{-23} \left(\frac{2~{\rm TeV}}
{\Lambda_{\rm ISS}}\right)\,.
$$
The actual observational bound of Raffelt and Weiss~\cite{raffelt}
for the coupling of a light pseudoscalar (e.g., the axion) to
electrons is $\alpha_a < 0.5 \times 10^{-26}$, assuming that the boson
is light enough to be produced in the star (by Compton scattering or
by bremsstrahlung), and assuming that it subsequently escapes.  This
constraint would rule out all PNGB models with $m_{\mathcal{P}}\lsim
10^4\hbox{\rm ---}10^5~{\rm eV}$, if $\Lambda_{\rm ISS}\lsim
4000$~TeV.\footnote{Although we have optimistically assumed that
$\Lambda_{\rm ISS}\sim 2$~TeV in this paper, in realistic models of
direct gauge mediation~\cite{direct}, $\Lambda_{\rm ISS}$ is no
larger than a few hundred TeV, well below the value needed to escape
the Raffelt and Weiss constraint.}

Thus, models with $N_C > 4$ (and in particular the Pentagon model)
are ruled out if the scale of meta-baryon number violation is $M_U$.
The model of \cite{pentabaryogen} is safe, as the scale of
penta-baryon number violation (in the intermediate range of
$10^8$---$10^{10}$ GeV) is considerably lower than $M_U$, in which
case the PNGB is too massive to be produced appreciably in the star.
The biggest unmet challenge for this latter model is to find an explanation for
this intermediate scale, with enough symmetry to ensure that it does
not contribute to dimension $6$ operators that violate ordinary
baryon number.  The most likely candidate for such a symmetry is a
discrete remnant of baryon number\footnote{Note that this discrete
symmetry does not prevent the generation of
the baryon asymmetry. Indeed, in the model of \cite{pentabaryogen}
electroweak baryon number violation, in combination with an
asymmetry in penta-baryon number, generates the cosmological baryon
asymmetry through spontaneous baryogenesis \cite{ck}.}
that is sufficient to prevent
proton decay (and prevent or suppress neutron-anti-neutron
conversion). Electroweak instantons conserve baryon number modulo
$6$ (in units where the baryon number of the proton is $1$).

\section{Conclusions}

We have investigated constraints on the PNGB of hidden sector baryon
number in SUSY breaking models based on the meta-stable states of
ISS.   We found no current laboratory constraints,
but constraints from the cooling of red giants require the PNGB mass
to be such that it cannot be produced in these stars. This puts
strong restrictions on the structure of the hidden sector gauge
group, if the irrelevant operators that break meta-baryon number
explicitly, are scaled by the unification scale.  Models in which
this scale is lower have to incorporate some mechanism for
suppressing dimension~$6$ operators that violate ordinary baryon
number.
\section*{Acknowledgments}
We would like to thank Linda Carpenter, Michael Dine, Guido Festuccia,
Michael Graesser, Michael Peskin and Yuri Shirman for
enlightening discussions.
This research was supported in part by DOE grant number
DE-FG03-92ER40689.

\end{document}